# Accessing MHz Operation at 2 V with Field-Effect Transistors Based on Printed Polymers on Plastic

*Andrea Perinot and Mario Caironi\**

Dr. A. Perinot, Dr. M. Caironi
Center for Nano Science and Technology@PoliMi, Istituto Italiano di Tecnologia, via Giovanni Pascoli 70/3, Milan, Italy
E-mail: mario.caironi@iit.it

Organic printed electronics is proving its suitability for the development of wearable, lightweight, distributed applications in combination with cost-effective production processes. Nonetheless, some necessary features for several envisioned disruptive mass-produced products are still lacking: among these radio-frequency (RF) communication capability, which requires high operational speed combined with low supply voltage in electronic devices processed on cheap plastic foils. Here, we demonstrate that high-frequency, low-voltage, polymer field-effect transistors can be fabricated on plastic with the sole use of a combination of scalable printing and digital laser-based techniques. These devices reach an operational frequency in excess of 1 MHz at the challengingly low bias voltage of 2 V, and exceed 14 MHz operation at 7 V. In addition, when integrated into a rectifying circuit, they can provide a DC voltage at an input frequency of 13.56 MHz, opening the way for the implementation of RF devices and tags with cost-effective production processes.

1. Introduction

Printed and solution-processed polymer Field-Effect Transistors (FETs) are considered among key enablers of flexible, wearable and portable electronics, and they hold the promise of low-cost access to many novel mass applications. This promise stemmed from the prospect of utilizing printing tools derived from well-established graphical arts technologies to deposit a variety of functional materials, which exhibit a suitable combination of mechanical and electronic properties.[1-5] Such vision has begun to show concrete potential through a variety of proof-of-concept demonstrations, among which programmable logic circuitry for flexible displays,[6] transferrable electronics for pharmaceutics,[7] healthcare sensors for brainwave detection[8] or pulse oximetry,[9] fully-printed washable electronics on fabrics[10] and imperceptible logic circuitry on ultra-thin substrates for intelligent electronic skin.[11] Nonetheless, several among the most desirable applications, such as Radio-Frequency (RF) tags and smart labels, driving circuitry for large-area high-resolution flexible displays and real-time sensor arrays, along with wireless sensors and sensors networks, require high-speed and low-voltage operation of the basic components of the circuits, in particular the transistor.[12, 13] Given the limited carrier mobility and the coarse resolution of printing tools, printed organic electronics has struggled to deliver the challenging performances required to enable wireless capabilities.[14-19] Alternative flexible electronics technologies with higher carrier mobilities are being developed to achieve high-frequency circuits, such as metal-oxide semiconductors, carbon nanotubes and 2-D materials, reaching in some cases very high *transition frequency $f_t$*,[20] the highest operative frequency of a transistor, in the order of the GHz or tens of GHz.[21, 22] However such high performances are either achieved by resorting to conventional micro- and nano-fabrication techniques (*i.e.* e-beam lithography, CVD, sputtering, thermal evaporation), or pose scaling and processing issues (placing of high quality monolayers of 2D materials,[23] alignment of CNTs,[24-27] process temperatures compatible with cheap plastic substrates for high-quality metal-oxide layers[28, 29]).

It is therefore highly desirable to further develop printed and flexible organic electronics in order to achieve high-frequency operation. While the possibility to obtain GHz organic transistors has only recently become argument of discussion,[30] progresses are being made in the range of near-field wireless communication. Several research groups in fact have proven the feasibility of operating organic transistors at frequencies in excess of 10 MHz,[31-36] but only a few works have achieved such frequencies by adopting printing techniques,[37] which are more easily upscalable and compatible with large-area processing. The maximum frequency drastically decreases for organic transistors on flexible substrates.[15, 17, 18] Moreover, since $f_t$ is proportional to the bias voltage, the requirement of low-voltage operation, at least below 10 V as necessary for portable, self-powered wireless electronics, further complicates the achievement of high operational frequency. Examples of $f_t$ beyond the MHz threshold have been shown through the adoption of conventional thermally-grown silicon dioxide dielectrics (20 MHz at a bias of 10 V[19]), ALD-deposited alumina (19 MHz at 10 V[33]) or hybrid metal-oxide/self-assembled ultrathin dielectrics (3.7 MHz at a bias of 3 V[38] and 1.5 MHz at 4 V[39]). There are instead scarce examples of fast devices where low-voltage is achieved through solution-processing. In this respect, ionic gating schemes through electrolyte dielectrics,[40, 41] which allow to achieve 1 V, or even lower, operational voltages, are not suitable because of their slow switching speed, in the tens of kHz range for the best reported cases,[42, 43] because of the slow movement of ions. Polymer dielectrics, in particular if based on low-k materials, are a suitable choice to attain ideal high-frequency operation of FETs, but the achievement of the high capacitance necessary for low-voltage operation is challenging. Demonstrations of MHz operation in FETs integrating low-k polymers are rare,[16] and high $f_t$ values at a voltage below 10 V of transistors of this kind on flexible substrate are yet to be shown.

In this work, we demonstrate that fully solution-processed, low-voltage polymer FETs, operating at MHz range, can be realized on flexible substrates with a combination of printing and direct-writing techniques. In particular, $f_t$ in excess of 1 MHz can be reached at an extremely low bias voltage of 2 V, a voltage comparable to electrolyte-gated devices. 14 MHz operation can be attained already at 7 V, therefore enabling near-field wireless communication, which we exemplify here with a voltage rectifier, one of the most relevant building blocks for 13.56 MHz RF tags. The combination of these performances, obtained through scalable processes on cheap flexible substrates, represents a step forward in the realization of mass-produced, distributed electronics with RF communication functionalities.

## 2. Results

To realize low-voltage, high-frequency polymer FETs we adopted a combination of femtosecond-laser sintering of high-resolution metal electrodes on plastic, a fast coating of uniaxially-aligned polymer semiconductor and a solution-processed high capacitance dielectric stack. We have previously adopted fs-laser sintering for the realization of high resolution electrodes and high-frequency polymer transistors on glass.[37] The same technique was also previously demonstrated to be compatible with plastic substrates for the fabrication of electrodes,[44, 45] semi-transparent grids[46] and transistors,[47-49] while it has not yet been adopted yet for high-frequency, printed electronics on plastic. Here we show that fs-laser sintering, thanks to highly controlled deposition of energy, allows the conversion of silver inks on plastic and the high spatial resolution patterning of electrodes suitable for high-performance fully solution-processed polymer transistors. Our simple two-steps fabrication process is sketched in **Figure 1a**: we first deposit via spin-coating a uniform thin film of an ink of silver nanoparticles, then a near-infrared fs-pulsed laser beam ($\lambda$ = 1030 nm, 67 MHz repetition rate) induces the local sintering of the nanoparticles, directly writing conductive patterns. Finally, the unprocessed areas are washed out with an organic solvent. In future implementations aiming at reducing the amount of waste materials, spin-coating may be replaced with a suitable printing technique (*e.g.* bar-coating or ink-jet) and the washed material can be recollected and recycled.

This approach allows to reach a maximum resolution of 1.5 µm for the patterning of conductive features at a scanning speed of 1 mm/s (current limit of our setup), with an impinging laser beam intensity of 1.9 mW µm$^{-2}$. With this method, we patterned silver source and drain electrodes with a width $L_{ov}$ of 1.7 µm on PEN to be used in FETs with a bottom-contact, top-gate architecture (Figure 1b). We realized an interdigitated structure in which the central finger is used as the source electrode, while the two external fingers are used as the drain. A photograph of the realized devices on a flexible PEN substrate is shown in Figure 1c. To proceed with the fabrication of OFETs, we followed the process illustrated in Figure 2a. We first modified the sintered Ag electrodes with a self-assembled monolayer of dimethylamino(benzenethiol) (DABT), which has been shown to improve charge injection for n-type organic semiconductors.[50] Then, we adopted the widely-studied semiconducting n-type co-polymer poly[N,N'-bis(2-octyldodecyl)-naphthalene-1,4,5,8-bis(dicarboximide)-2,6-diyl]-alt-5,5'-(2,2'-bithiophene), P(NDI2OD-T2) and deposited a thin layer of such material via bar-coating, which is a simple and fast method to induce the directional alignment of the polymer chains along the coating direction,[51] yielding a film with optimized charge mobility over large area.[14] We then deposited via spin-coating a 150-nm-thick multi-layered polymer dielectric stack, composed of an ultra-thin crossed-linked low-k polymer and a top high-k polymer, achieving an areal capacitance $C_{diel}$ of 39 nF cm$^{-2}$. We finally inkjet-printed PEDOT:PSS gate electrodes on top of the dielectric. We fabricated FETs (Figure 2b) with a varying channel length $L$, ranging from 1 to 17.5 µm, with constant channel width $W$ of 800 µm.

The transfer curve for a FET with the shortest channel length of 1 µm (Figure 2c) highlights the correct operation and turn-on of the device at a gate voltage as low as 5 V, both in the linear ($V_d = 1$ V) and in the saturation ($V_d = 5$ V) regime, with no appreciable hysteresis. In addition, we highlight the extremely low gate leakage current that is more than 4 orders of magnitude lower than the device ON current in both operation regimes. Similarly, analogous electrical performance and low gate leakage current are achieved in the devices with longer channel length (Figure S1a-c). The drain current scales correctly with the channel length (Figure 2e, S2a-c), while the output curve, although not achieving clear saturation differently from the longer channel devices (Figure S3 a-c) and likely owing to the onset of short-channel effects,[52] both confirms the correct operation of our downscaled device and suggests that a good charge injection performance is achieved in our architecture (Figure 2d).

We calculated the apparent electron mobility ($\mu_{app}$) for our devices as a function of gate voltage form the derivative of the transfer curves according to the gradual channel approximation model. In the device with $L = 1$ µm (Figure 2f), $\mu_{app}$ peaks at a low $V_g$ of 1.8 V (2.2 V) in the linear (saturation) regime, then slightly rolls off with $V_g$, an effect related to the influence of charge injection limitations. For our case, we simply report in Table 1 the calculated value for the mobility in the saturation regime at maximum $V_g$, being one of the relevant bias point for the FET and for the rectifier circuit presented later in the text. At this bias point, $\mu_{app}$ ranges in the interval from 0.15 to 0.3 cm$^2$ V$^{-1}$ s$^{-1}$ when $L$ is shortened from 17.5 to 1 µm. Such an increasing trend as the channel length is reduced is an effect stemming from the increase of the lateral electric field across the channel region, which affects charge injection and/or charge transport in organic FETs.[53-55] In Table 1, alongside with the calculated mobility, we report the *measurement reliability factor r* as suggested in Choi et al.,[56] which in our cases is always above 100 %, highlighting the presence of a very mild "kink effect".

We extracted the contact resistance of the shortest-channel device ($L = 1$ µm) with the differential method,[57] which yielded a width-normalized contact resistance $R_CW = 1015$ Ωcm at $V_g = 5$ V in the linear regime. The latter is a very small value in the context of printed polymer FETs, has rarely been achieved with the sole use of solution-based approaches and with low-voltage operation[58, 59] and has a paramount role in the achievement of good electrical behavior of our downscaled FETs. In particular, $R_CW$ is below the value of the width-normalized channel resistance $R_{ch}W = V_dW/I_d - R_cW = 1369$ Ωcm at the highest gate bias of $V_g = 5$ V (and $V_d = 1$ V). Despite $R_CW$ below 1000 Ωcm have been demonstrated for organic transistors,[60-66] these examples either adopted evaporation of dopants/electrodes, or device biasing in excess of 10 V or the use of electrolyte-gating schemes, which are not desirable for high-frequency, low-voltage, all-solution-processed devices.

We show in Figure S4a-b the superimposed transfer curves of 5 different devices each for $L = 1$ µm and $L = 1.5$ µm cases, and in Figure S5 we plot the superimposed trends of the extracted mobility versus gate voltage for the devices with the shortest channel ($L = 1$ µm), highlighting that our results are reproducible even at these challengingly short, micron-sized, channel lengths. In Table S1 we also report the mean and standard deviation of some selected figures of merit (i.e. the mobility at maximum bias and the parameters $\mu_0$, $\gamma$ and $V_T$ as defined in *Natali et al.* [57]) for the set of 5 devices with $L = 1$ µm. Once again, the reproducibility of our devices is highlighted especially in terms of intrinsic mobility $\mu_0$ and in terms of the threshold voltage $V_T$, for which the standard deviation is respectively 1.4 % and 5 % of the mean value. We also measured the transfer curve of a device ($L = 1$ µm) after 7 months from fabrication and storage in nitrogen atmosphere, highlighting how only a small loss in performance is detected (Figure S6a). We then kept such transistor in the ON state for 150 minutes by applying $V_g = V_d = 5$ V and measured its transfer curves every 30 minutes (Figure S6b), detecting a good stability of the device upon operational stress.

In order to evidence the benefit of downscaled features on the frequency response of the fabricated FETs, we characterized the AC performance of our devices in terms of transition frequency $f_t$, which depends on the electrical parameters of the transistor according to:[20]

$$f_t = \frac{g_m}{2\pi \ (C_{gs} + C_{gd})}$$

where $g_m$ is the device transconductance and $C_{gs}$, $C_{gd}$ are the gate/source and gate/drain capacitances respectively. Additional details on the measurement method can be found in [37]. In Figure 3a we show the measured $C_{gs}$ and $C_{gd}$ versus bias voltage in the saturation regime ($V_{gs} = V_{ds}$), and compared them to the theoretically expected values for our layout, according to:[67]

$$C_{gs} = C_{diel}\left(W\frac{2}{3}L + W_{ov}L_{ov}\right)(1 + \alpha_{f,gs})$$

$$C_{gd} = 2C_{diel}W_{ov}L_{ov}(1 + \alpha_{f,gd})$$

where $W_{ov}$ = 500 µm is the total width of the overlap between gate and bottom electrodes, which exceeds the actual channel width $W$ (see the scheme reported in Figure S7), while the correction factors $\alpha_{f,gs}$ and $\alpha_{f,gd}$ are introduced to account for the fringing capacitance between the source/drain electrodes and the wide top gate. We calculated such factors using a formula proposed by *Elmasry*,[68] yielding $\alpha_{f,gs}$ = 12.41 % and $\alpha_{f,gd}$ = 21.05 %, which is in perfect agreement with our measured data. In the case of $C_{gd}$, the measured values are correctly constant at ~0.8 pF up to a bias voltage of 5 V, while $C_{gs}$ exhibits a slightly increasing trend with the bias voltage, which we attribute to parasitic accumulation of additional charge outside the channel active area. This effect has already been explained for structures of this kind,[69] and originates from the fact that the semiconductor layer is not patterned, covering the whole substrate. At bias voltages > 5 V, the higher electric fields are producing a further increase in capacitance and a relatively small deviation from the predicted values.

The measured $g_m$ per unit width versus bias voltage is shown in Figure 3b. We correctly identify a linear increase of $g_m$ from $V_{gs}$ = 1 V to $V_{gs}$ = 5 V, where the transconductance varies from 0.081 to 0.58 mS cm$^{-1}$. The latter value is very close to the theoretical one extracted from the slope of the transfer curves, which yields $\frac{g_m}{W} = \frac{\mu C_{diel}}{L}(V_{gs} - V_t) = 0.53 \frac{mA}{Vcm}$. For bias voltages in excess of 5 V, $g_m$ exhibits a superlinear increase with $V_{gs}$, an effect that we attribute to the high lateral electric field insisting across the channel region. The measurements for $g_m$, $C_{gs}$ and $C_{gd}$ can be combined to identify $f_t$ as in Figure 3c, where we show such measurement for a device with $L$ = 1 µm and a bias voltage of 7 V (top panel) and 2 V (bottom panel). The measurement is limited at a frequency of 2 MHz due to setup constraints. Remarkably, we can measure an $f_t$ = 1.6 MHz for a bias voltage of only 2 V. In Figure 3d we show the trend of the measured $f_t$ for the same device versus bias voltage, highlighting how this figure of merit essentially follows the behavior of $g_m$. In the case of a bias voltage of 7 V, we can extrapolate a high transition frequency of 10.4 MHz.

To assess the relevance of the performance achieved by our devices and also to favor the identification of effective strategies to improve even further $f_t$, we have calculated the maximum achievable transition frequency with respect to the variation of our FET parameters, using the modeling recently proposed by *H. Klauk*.[30] With an $R_CW$ of a value equal to the one we measured, neither a further reduction of $L$ nor an increase of $\mu_0$ would yield a very significant increase in $f_t$ (Figure S8a). At such fixed contact resistance, an increase in $f_t$ would only be obtainable with the reduction of $L_{ov}$. However, even in the extreme and unrealistic case of $L_{ov} = 0$, $f_t$ would be limited to a value below 100 MHz at best (Figure S8b). These calculations underline that we have already obtained, with the physical parameters characterizing our devices, an AC performance that is close to the highest possible. The reduction of the contact resistance thus constitutes the main route for accessing higher $f_t$ in the future, as it is the key point in order to achieve a regime where the modification of the other parameters is effective (Figure S8c).

We have also measured the reproducibility of our devices in terms of transition frequency after 7 months from fabrication. In Figure S9 we show the measured curves for 4 devices with $L = 1$ µm at a bias voltage of 5 V, which highlight good reproducibility for $f_t$ with a mean value of 6.65 MHz and a standard deviation of 2.12 MHz.

Overall, our best $f_t$ in the measurement set yielded a value of 14.4 MHz at a bias voltage of 7 V (Figure S10).

We have also compared our achieved performance to the previous literature for organic high-frequency transistors on plastic, for which the transition frequency was explicitly measured and reported. Since $f_t$ is directly proportional to the applied bias voltage, in our comparison we decouple such effect by defining the voltage-normalized transition frequency $f_t/V_{bias}$, where $V_{bias}$ is the maximum voltage applied to the transistor electrodes during $f_t$ measurement.

In this way the voltage-normalized transition frequency encompasses the achieved performance exclusively in terms of effective charge transport properties and geometrical resolution. We report in Table 2 the calculated $f_t/V_{bias}$ of our FETs in comparison with other representative examples of high-frequency ($f_t > 1$ MHz) organic devices on flexible substrate. Our FETs achieve the best performance in terms of $f_t/V_{bias}$ in the field of organic transistors on plastic, with a tenfold improvement compared to previous demonstrations.

While $f_t$ is the relevant figure of merit for the assessment of the FET maximum operational speed for integration in analog circuitry (*e.g.* amplifiers), we additionally assessed the performance of our FETs upon large-signal operation, which is relevant for the implementation of RF applications such as wireless data communication. To this goal, we realized a simple rectifier circuit (Figure 4b, inset) and operated it up to 15 MHz, above the frequencies of interest for NFC. In Figure 4a we show the rectified output (red line) on an oscilloscope corresponding to an oscillating input signal with an amplitude of 8 V (black line). We measure a -3 dB frequency of 8.7 MHz, with respect to the amplitude recorded at 100 kHz (Figure S11). For a reduced signal amplitude of 5 V, the -3 dB frequency is as high as 6.4 MHz already (Figure 4b), and 1 V can still be read in DC at 15 MHz (Figure S12) proving the applicability of our approach for the implementation of RF tags.

We also verified the stability of the output of our rectifier upon constant operation for 150 minutes at an input voltage amplitude of 5 V and at an input frequency of 15 MHz (Figure S13). The device was measured after 7 months from fabrication. Our rectifier shows stable performance over 2.5 hours of continuous operation even at the high operational frequency of 15 MHz.

To better understand the measured performance of our rectifier we developed a simple model, based on the charge balance of the smoothing capacitor (see Figure 4b, inset), for the behavior of the output voltage versus input voltage frequency (see Supplementary Information). We are able to predict with good approximation the frequency behavior of the output voltage (Figure 4b, solid line), correctly identifying the frequency cutoff. Moreover, the model allows us to identify the main factors determining the output/input voltage ratio without affecting the rectifier cutoff frequency: in order to achieve a high ratio, it is critical to limit the OFF current in the transistor and to adopt a channel width large enough to feed the load resistance. The cutoff frequency is instead determined by the same parameters as the $f_t$ (*i.e. L, μ, $V_{bias}$*) and can be set independently.

## 3. Conclusions

We have demonstrated the realization of high-frequency, low-voltage, polymer FETs on flexible substrates using only printing techniques and digital laser patterning. The proposed FETs feature a transition frequency in excess of 1 MHz at a bias voltage as low as 2 V, and can operate at a frequency in excess of 10 MHz at a bias voltage of 7 V, which is, to the best of our knowledge, the highest for the case of printed polymers integrating direct-written electrodes, on flexible substrates. Such performance allows the reported devices to operate at supply voltages comparable to electrolyte gated organic transistors, while they can be modulated at 2-3 orders of magnitude higher frequencies. Moreover, such operational bias voltage enables the implementation of stand-alone electronic devices, being fully compatible with thin film batteries or energy harvesters, such as plastic photovoltaic modules.[70] In addition, these devices achieve the best figure to date, by an order of magnitude, in terms of voltage-normalized transition frequency (2.06 MHz V$^{-1}$). Finally, we have integrated these devices into a rectifying circuit that can provide voltage rectification up to 15 MHz, thus being compatible with NFC wireless communication at 13.56 MHz. The demonstration of such performance and its achievement with the sole use of solution-based processing, printing techniques and laser patterning, identifies a feasible route towards low-cost, mass-scale production of organic devices for RF applications.

## 4. Experimental Section

*General*: The Ag-nanoparticles ink (NPS-JL) was purchased from Harima Chemicals, 4-(Dimethylamino)benzenethiol (DABT) was purchased from TCI Chemicals, P(NDI2OD-T2) was purchased from Polyera (Mn = 35.3 kDa; PDI = 1.8 (GPC); Elemental analysis: C: 75.21, H: 8.73, N: 2.87 (Theoretical: C:75.26, H:8.96, N: 2.83)), PEDOT:PSS (Clevios PJ700) was purchased from Haereus. The substrates consist of 125-µm-thick polyethylene naphthalate foils (Teonex Q65FA) purchased from DuPont TeijinFilms.

*Femtosecond laser sintering setup:* The laser setup consists of a commercial laser source ( LightConversion PHAROS, based on Yb:KGW as active medium) which generates ~80 fs-long laser pulses with a repetition rate of 67 MHz, $\lambda$ = 1030 nm and maximum output power of 2 W. Before reaching the sample, the beam is conditioned through an optical path which includes a software-controlled attenuator and a focalizing objective (Mitutoyo) lens whose magnifying power can be selected between 20X, 50X or 100X. The sample is positioned on a software-controlled moving stage (Aerotech ABL1000) capable of a maximum resolution of 0.5 nm and a maximum speed of 300 mm s$^{-1}$.

*Contacts fabrication:* The PEN substrates are used as-is on the pristine side (without adhesion-promoting treatment) and a 4-nm-thick AlOx layer is deposited via thermal evaporation of Al and exposure to atmosphere. The Ag-nanoparticles ink is spun at 7000 rpm for 5 min, yielding a 70-nm-thick layer. The laser-sintering step is then performed using 1.9 mW µm$^{-2}$ beam power, 50X optics, 0.1 mm s$^{-1}$ scanning speed. After laser processing, the samples are thoroughly washed with o-Xylene and isopropanol and finally dried with a nitrogen flux.

*Organic FETs fabrication:* The samples with the desired laser-sintered patterns for the FET bottom electrodes are Ar-plasma etched with a power of 100 W for 4 min. A solution of 17 µl of DABT in 12 ml of isopropanol is prepared, the samples are immersed in it for 15 minutes and the rinsed with abundant isopropanol. A solution of P(NDI2OD-T2) in toluene (5 g l$^{-1}$) is then deposited via bar-coating on the samples as described in [14], using a bar designed to yield a 8-µm-thick wet layer. The samples are then annealed in a nitrogen atmosphere for 20 min at 120 °C. The dielectric is a double layer stack[71] composed of a very thin layer of a low-k polymer, in contact with the semiconductor, and a top high-k polymer, for a total thickness of about 200 nm and an areal capacitance of 39 nF/cm$^2$. Finally, PEDOT:PSS is inkjet-printed with a Fujifilm Dimatix DMP-2831 to pattern the gate electrodes and the samples are annealed for 8 h in nitrogen atmosphere before measurement.

*Rectifier fabrication:* The rectifier circuit is realized by connecting one of the realized transistor, biased in a transdiode configuration, to a discrete capacitor of 0.15 µF and to an oscilloscope with an input resistance of 1 MΩ using external point-contact probes.

*Electrical characterization:* The devices are measured in nitrogen atmosphere. Static characterization is performed via a Keysight B1500A Semiconductor Parameter Analyzer. Frequency performance was measured using a custom setup which includes a Keysight ENA E5061B Vector Network Analyzer and an Agilent B2912A Source Meter. The apparent mobility $\mu_{app}$ is extracted from the transfer characteristic according to $\mu_{app} = \frac{2L}{WC_{diel}} \left( \frac{\partial \sqrt{I_d}}{\partial V_g} \right)^2$ for the saturation regime and to $\mu_{app} = \frac{L}{V_d W C_{diel}} \frac{\partial I_d}{\partial V_g}$ for the linear regime.

**Supporting Information**

Supporting Information is available from the Wiley Online Library or from the author.

**Acknowledgements**

The authors are thankful to L. Criante and his team for the support with the femtosecond laser machining setup. This work has been financially supported by the European Research Council (ERC) under the European Union's Horizon 2020 research and innovation programme 'HEROIC', grant agreement 638059.

**Figure 1.** a) Laser sintering processing steps for the fabrication of conductive electrodes on plastic, b) optical micrograph of a single device comprising active area and contact pads (magnification: layout of the source and drain electrodes), c) picture of the final realized devices on plastic PEN substrate.

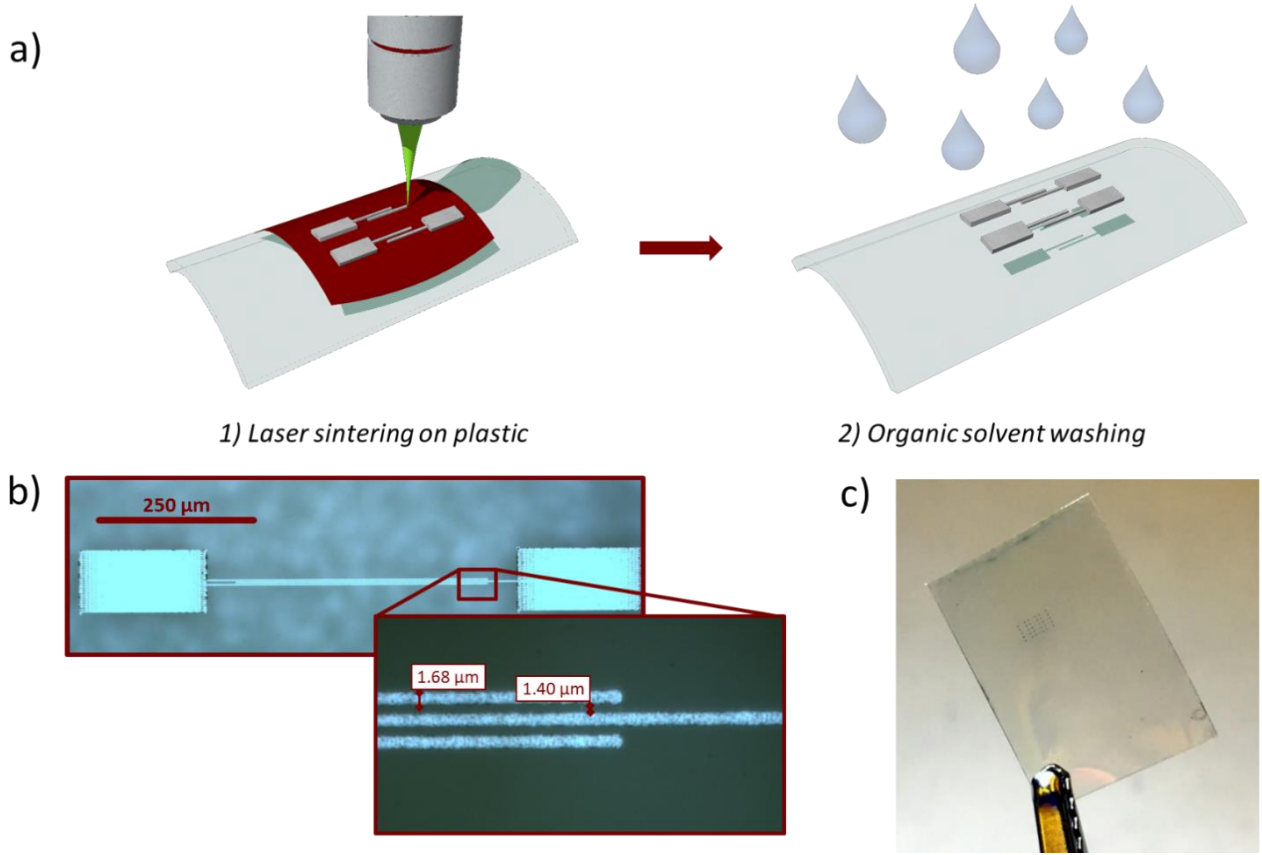

**Figure 2**. a) Process flow for the fabrication of polymer FETs on plastic with solution-based techniques, b) 3D view of the final device stack, c) transfer curve and d) output curve for a realized FET with $L = 1$ µm, e) drain current at $V_g = 5$ V for the FETs and for the realized channel lengths, f) calculated apparent charge mobility for the device with $L = 1$ µm.

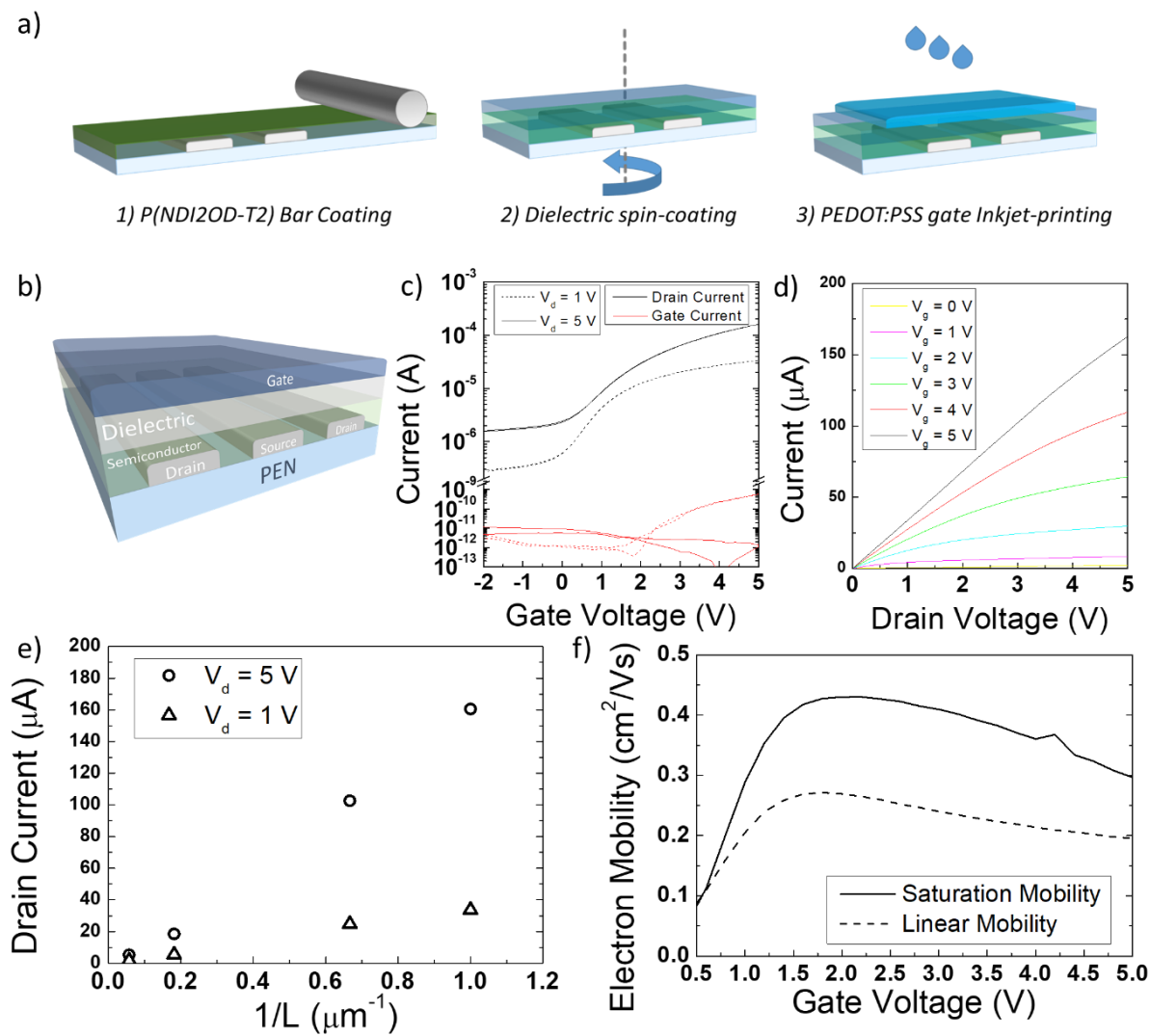

**Figure 3.** a) Measured and calculated data for the gate-source and gate-drain capacitances of a FET with $L = 1$ µm versus gate-source bias voltage, b) measured transconductance of the same device versus $V_{gs}$ (the linear fit is a guide to the eye), c) combined measurement of the transconductance and gate capacitance of the same FET for $f_t$ extraction (top panel: 7 V bias, bottom panel: 2 V bias), d) extracted transition frequency for the same FET versus gate-source bias voltage.

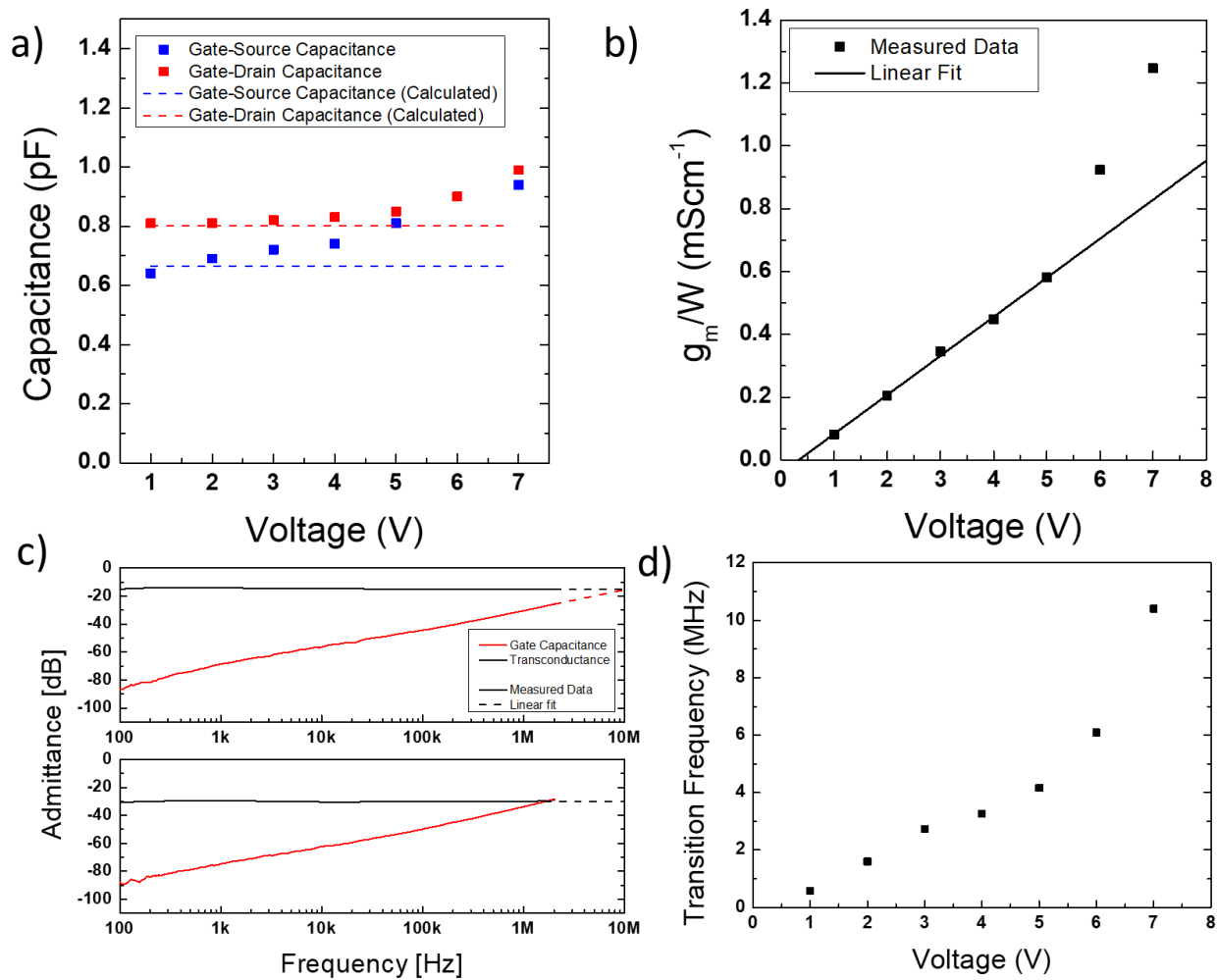

**Figure 4.** a) Input and output voltage waveforms for a rectifier based on a realized FET with $L = 1$ µm at an input voltage frequency of 100 kHz (top panel) and of 10 MHz (bottom panel). b) Measured and simulated output voltage of the same rectifier versus frequency at an input voltage amplitude of 5 V (inset: rectifier circuit).

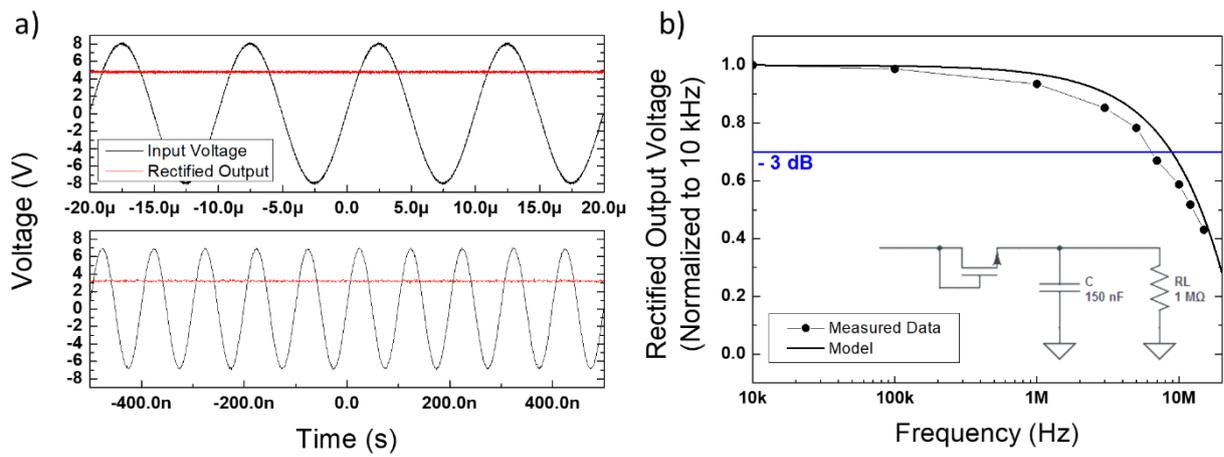

**Table 1.** Measured effective charge mobility and corresponding reliability factor for the realized FETs for the different channel lengths

| Channel Length [µm] | Apparent Mobility at $V_g$ = 5 V [cm$^2$ V$^{-1}$ s$^{-1}$] | Reliability factor $r$ |
|---|---|---|
| 1 | 0.30 | 107% |
| 1.5 | 0.29 | 109% |
| 5.5 | 0.16 | 128% |
| 17.5 | 0.14 | 134% |

**Table 2.** Achieved voltage-normalized transition frequency $f_t/V_{bias}$ for relevant works on organic, high-frequency FETs on flexible substrate.

| $f_t$ [MHz] | $V_{bias}$ [V] | $f_t/V_{bias}$ [MHz V$^{-1}$] | Fabrication techniques | Ref. |
|---|---|---|---|---|
| 14.4 | 7 | 2.06 | Laser sintering, bar-coating, spin-coating, inkjet. | This work |
| 1.6 | 2 | 0.80 | Laser sintering, bar-coating, spin-coating, inkjet | This work |
| 1.6 | 8 | 0.20 | Inkjet, spin-coating | [16] |
| 4.9 | 30 | 0.16 | Laser ablation, spin-coating, inkjet | [18] |
| 1.92 | 15 | 0.13 | Gravure, inkjet | [15] |
| 3.3 | 30 | 0.11 | NIL[a)], evaporation, inkjet, gravure | [17] |

[a)] **NIL: Nano-Imprint Lithography**